\documentclass[12pt]{article}
\usepackage{epsf}

\usepackage{pazha}
\tightenlines

\def\*{$^{*}$}
\def\a{$^{\mbox{\small a}}$}
\def\b{$^{\mbox{\small b}}$}
\def\c{$^{\mbox{\small c}}$}
\def\d{$^{\mbox{\small d}}$}
\def\e{$^{\mbox{\small e}}$}
\def\f{$^{\mbox{\small f}}$}
\def\g{$^{\mbox{\small g}}$}
\def\h{$^{\mbox{\small h}}$}
\def\i{$^{\mbox{\small i}}$}

\def\etal{{et~al.}}
\sloppypar

\begin{document}
\baselineskip 18pt
\noindent
{\it to be published in Astronomy Letters, 2016, v.\,42,
  pp.\,69-81}\\ [1cm]

\title{\bf X-RAY NOVA MAXI\,J1828--249. EVOLUTION OF THE
  BROADBAND SPECTRUM DURING ITS 2013--2014 OUTBURST} 

\author{\bf 
S.A.~Grebenev\affilmark{1*}, A.V.~Prosvetov\affilmark{1}, 
R.A.~Burenin \affilmark{1},\\ R.A.~Krivonos\affilmark{1}, 
and A.V.~Mescheryakov\affilmark{1,2}}   

\affil{
$^1${\it Space Research Institute, Russian Academy of Sciences,
    Profsoyuznaya ul.\\ 84/32, Moscow, 117997 Russia \makebox[74mm]{}}\\
$\!\!\!\!\!^2${\it Kazan Federal University, Kremlevskaja ul. 18, Kazan, 420008 Russia\makebox[6mm]{}}\\}

\vspace{2mm}
\received{June 5, 2014}

\sloppypar 
\vspace{2mm}
\noindent
Based on data from the SWIFT, INTEGRAL, MAXI/ISS orbital
observatories, and the ground-based RTT-150 telescope, we have
investigated the broadband (from the optical to the hard X-ray
bands) spectrum of the X-ray nova MAXI\,J1828--249 and its
evolution during the outburst of the source in 2013--2014. The
optical and infrared emissions from the nova are shown to be
largely determined by the extension of the power-law component
responsible for the hard X-ray emission. The contribution from
the outer cold regions of the accretion disk, even if the X-ray
heating of its surface is taken into account, turns out to be
moderate during the source's ``high'' state (when a soft blackbody
emission component is observed in the X-ray spectrum) and is
virtually absent during its ``low'' (``hard'') state. This result
suggests that much of the optical and infrared emissions from
such systems originates in the same region of main energy
release where their hard X-ray emission is formed. This can be
the Compton or synchro-Compton radiation from a high-temperature
plasma in the central accretion disk region puffed up by
instabilities, the synchrotron radiation from a hot corona above
the disk, or the synchrotron radiation from its relativistic
jets.  

\noindent
{\bf DOI:} 10.1134/S1063773716020031


\vfill
\noindent\rule{8cm}{1pt}\\
{$^*$ E-mail: $<$sergei@hea.iki.rssi.ru$>$}

\clearpage
\section*{INTRODUCTION}
\noindent 
The outburst of the previously unknown X-ray transient
MAXI\,J1828--249 at $\sim1\fdg5$ from the accreting pulsar
4U\,1826-24 was detected by the GSC/MAXI instrument on October
15, 2013, at 21\uh55\um\ UT (Nakahira et al. 2013). By the time
of the discovery, the photon flux from the source reached
$\sim90$ mCrab in the 2--10 keV energy band; the spectrum was
soft, a blackbody one. The position of the transient in the sky,
initially determined with an accuracy of $\sim 0$\fdg$3$, was
improved during its observation with the IBIS/ISGRI telescope of
the INTEGRAL observatory (Filippova et al. 2013) and,
subsequently, with the XRT and UVOT telescopes of the SWIFT
observatory (Kennea et al. 2013a, 2013b): $R.A. (\mbox{J}2000) =
18$\uh$28$\um$58$\fsec$07$ and $Dec. (\mbox{J}2000) =
-25$\deg$01$\arcmin$45$\farcs$88$ (the uncertainty is
$0$\farcs$03$). The source turned out to be faint in the
ultraviolet, $M2\simeq18.6$. Given the very moderate absorption
measured in its X-ray spectrum, $N_{\rm H}\sim 2\times
10^{21}\ \mbox{\rm cm}^{-2}$, it can be attributed to low-mass
X-ray binaries. The photon flux from the source in the 20--80
keV energy band was $\sim45$ mCrab, while its spectrum was hard,
a power law with a photon index $\alpha\sim 1.7$ (Filippova et
al. 2013). This result can be brought into agreement with the
measurements of Nakahira et al. (2013) at low energies only by
assuming that the source has a two-component spectrum similar to
the spectra of X-ray novae\footnote{Low-mass X-ray binaries
  containing a black hole that usually do not emit X-rays but
  occasionally flare up due to nonstationary accretion.} in
their bright state (see, e.g., Sunyaev et al. 1988, 1991; Tanaka
and Shibazaki 1996; Grebenev et al. 1997; Remillard and
McClintock 2006; Cherepashchuk 2013). The subsequent
observations performed on October 17--18 with the SWIFT and
INTEGRAL observatories (Krivonos and Tsygankov 2013) showed that
the source's spectrum rapidly softened ($\alpha\ga2$), which is
also typical of X-ray novae on their way to maximum light. The
radio observations made on October 18 with the Australian ATCA
telescope revealed no emission from the source (Miller-Jones
2013). Note that the absorption measured in the source's
spectrum corresponded to the expected Galactic one in this
direction, $N_{\rm H}\simeq (1.7\pm0.2)\times10^{21}\ \mbox{\rm
  cm}^{-2}$ (Kalberla et al. 2005).

Thus, already the first days of observations showed
that an X-ray nova flared up in the Galactic center
region, and we had a unique chance to investigate the
properties of yet another representative of the so far
small population of these interesting objects and to
test the existing theoretical models of disk accretion
onto a black hole in a binary system. In this paper,
we present the results of our monitoring of the nova
outburst by the INTEGRAL (Winkler et al. 2003),
MAXI (Matsuoka et al. 2009), and SWIFT (Gehrels
et al. 2004) international astrophysical observatories.
We use the publicly accessible data and the INTEGRAL
data obtained within the Russian quota of
observing time.

Of particular interest to us was the evolution of the broadband
spectrum for this nova and primarily the properties of its
optical, infrared (OIR), and ultraviolet (UV) emissions. It is
generally believed that such an emission appears in low-mass
X-ray binaries due to the irradiation and heating of the outer
cold accretion disk by X-ray photons from its hot central zone
(Lyuty and Sunyaev 1976). However, recent studies of the
broadband spectra for several X-ray novae, XTE\,J1118+480 (Chaty
et al. 2003), MAXI\,J1836--194 (Grebenev et al. 2013), and
SWIFT\,J174510.8--26241 (Grebenev et al. 2014), in their hard
spectral state have shown that the OIR and UV emissions from
these sources are an extension of the power law observed in the
X-ray band and contain no clear evidence of a thermal emission
component from the outer disk regions (see also Poutanen and
Veledina 2014). Observations of the X-ray nova MAXI\,J1828--249
can give additional information for investigating this question.

\section*{INSTRUMENTS AND DATA ANALYSIS}

The data from the ISGRI detector (Lebrun et al.  2003) of the
IBIS gamma-ray telescope (Ubertini et al. 2003) onboard the
INTEGRAL observatory obtained within the Russian quota of
observing time and in public access programs were used to study
the outburst of this source. This detector is sensitive in the
energy range 18--200 keV. Unfortunately, during the observations
being discussed, the source never fell within the field of view
of the INTEGRAL JEM-X monitor (Lund et al. 2003), which is
narrower (with a diameter of 13\fdg2) than that of the IBIS
telescope ($30\deg \times 30 \deg$).

At the SWIFT observatory, we used data from
the BAT gamma-ray telescope sensitive in the energy
range 15--150 keV with a field of view of 1.4 sr
(Barthelmy et al. 2005), the XRT telescope sensitive
in the energy range 0.2--10 keV (Burrows et al.
2005), and the UVOT optical and ultraviolet telescope
sensitive in the range 170--600 nm (Roming
et al. 2005). To obtain the sky images and to study
the properties of individual sources, coded-aperture
masks are used in the IBIS and BAT telescopes
and grazing-incidence mirrors are used in the XRT
telescope. We analyzed the data from all SWIFT
telescopes using the standard data processing software
packages (see also Evans et al. 2010 and
references therein). The IBIS/ISGRI data were processed
using the software developed at the Space Research
Institute of the Russian Academy of Sciences
(see, e.g., Revnivtsev et al. 2004; Krivonos et al.
2010). Our spectral analysis was performed with the
NASA/HEASARC/XSPEC software package (Arnaud
et al. 1996).

In the MAXI experiment onboard the International
Space Station (ISS), we used data from the
Gas Slit Camera (GSC) sensitive in the energy
range \mbox{2--30} keV and scanning the entire sky every
90 min (Mihara et al. 2011). We used only the
automatically processed light curves from the site
{\sc maxi.riken.jp/top} averaged on a scale of one day.

The optical and infrared observations of the source
were performed with the Russian-Turkish 1.5-m
telescope (RTT-150) on November 9--12, 2013.
The detector was the TFOSC spectrometer, which
operated for these measurements in the regime of a
photometer in the {\it v, b, z, i, r,} and {\it g} filters, each
with an exposure time of 60 s. We performed the
photometric calibration using standard stars and
processed the observations using the IRAF software
package.

\section*{RESULTS}

\begin{figure}[tp]

\begin{minipage}[t]{0.5\textwidth}
\epsfxsize=1.0\textwidth
\epsffile{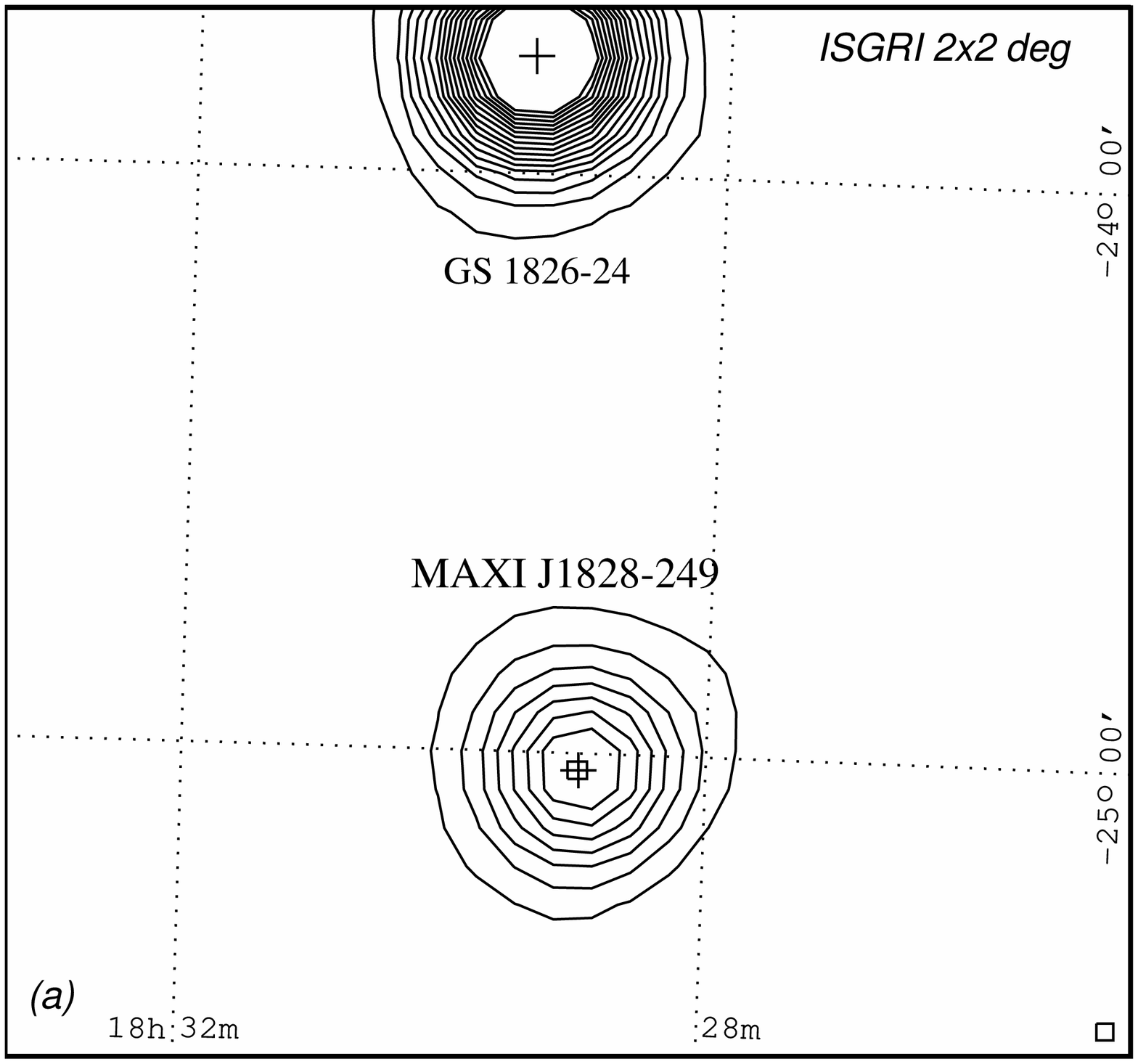}
\end{minipage}\,\begin{minipage}[t]{0.5\textwidth}
\epsfxsize=1.0\textwidth
\epsffile{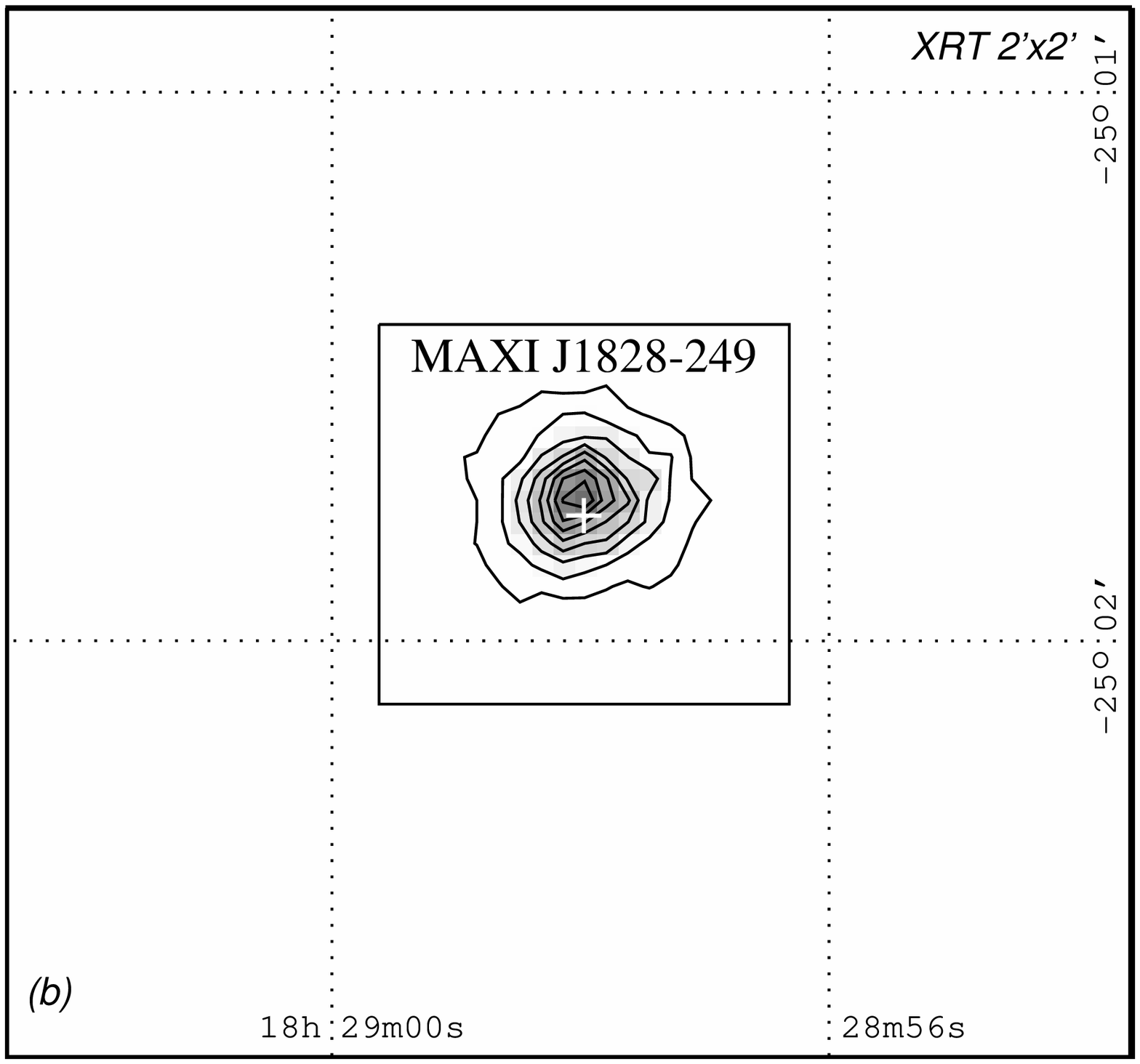}
\end{minipage}\\
\begin{minipage}[t]{0.5\textwidth}
\epsfxsize=\textwidth
\epsffile{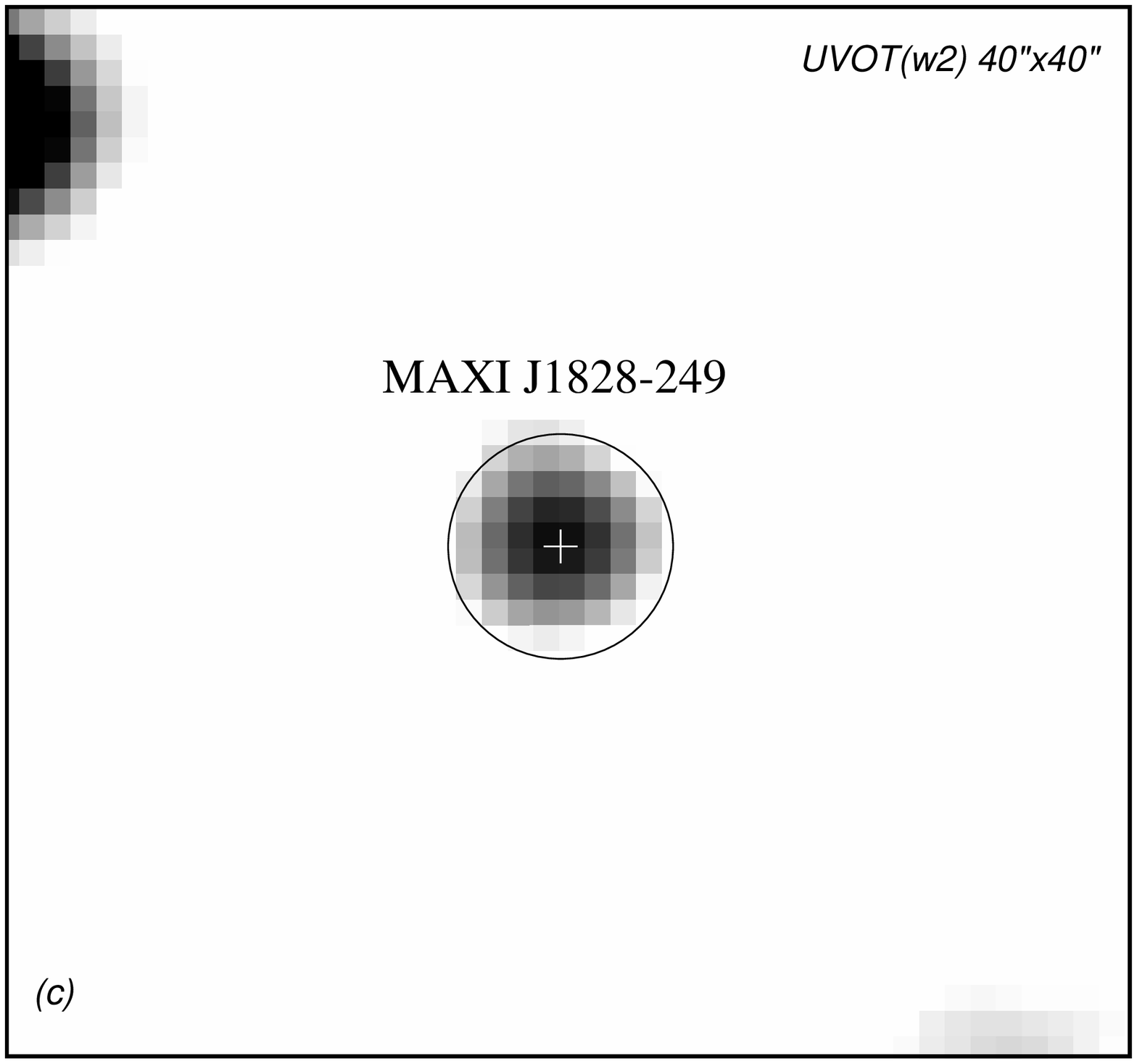}
\end{minipage}\,\begin{minipage}[t]{0.502\textwidth}
\epsfxsize=\textwidth
\epsffile{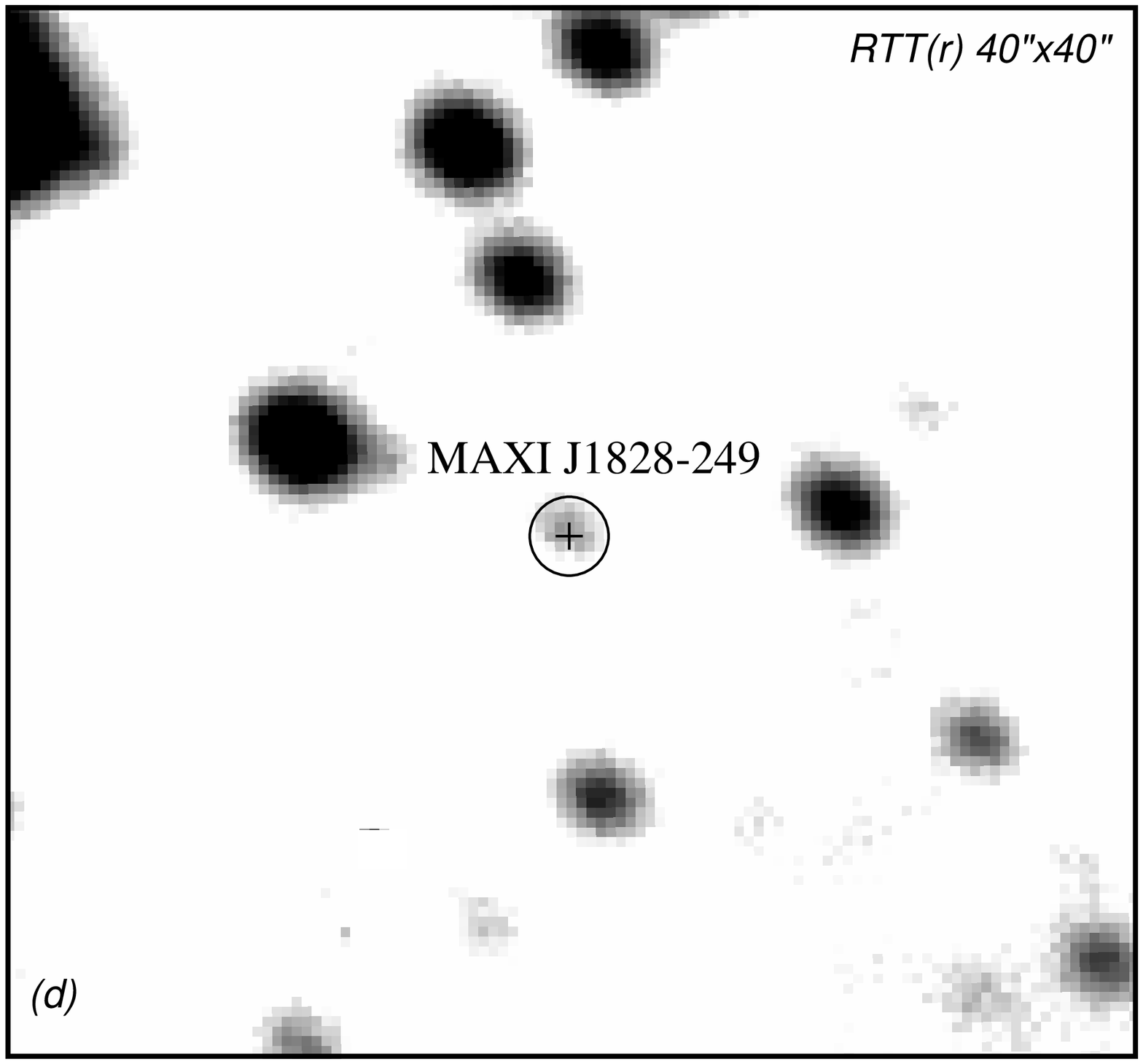}
\end{minipage}

\caption{\rm Images of the sky region near MAXI\,J1828--249
  obtained with the IBIS/ISGRI telescope of the INTEGRAL
  observatory (20--60 keV), XRT (0.3--10 keV) and UVOT ({W2}
  filter) telescopes of the SWIFT observatory, and the 1.5-m
  RTT-150 telescope ({r} filter). The $S/N$ map is shown in
  the case of IBIS/ISGRI, and the intensity map is shown in the
  remaining cases. The images are given in order of improving
  angular resolution of the telescopes: 12\arcmin, 5\arcsec,
  2\farcs5, and $\sim2$\arcsec\ (seeing). Their size decreases
  accordingly: 2\deg$\times$2\deg, 2\arcmin$\times$2\arcmin, and
  40\arcsec$\times$40\arcsec\ (for the UV and IR images). The
  region in the UV and IR images is marked by the square in the
  XRT image; in turn, the XRT region is marked by the small
  square in the ISGRI image (for clarity, its size is also shown
  in the lower right corner of this image). The radius of the
  circumferences in the UV and IR images is 4\arcsec\ and
  2\farcs5, respectively.}
\end{figure}
The IBIS/ISGRI/INTEGRAL, XRT and UVOT/SWIFT, and RTT-150 images
of the sky region in which MAXI\,J1828--249 flared up are
presented in Fig.\,1. The images are given in order
($a\rightarrow b\rightarrow c\rightarrow d$) of improving
angular resolution of the telescopes (from 12\arcmin\ for
IBIS/ISGRI to $\sim2$\arcsec\ for RTT-150; in the latter case,
the average seeing is specified). The figure shows an ordinary
path that has to be traversed from the detection of a new source
by a wide-field X-ray telescope to its highly accurate optical
localization. Owing to the moderate absorption and confident
detection of MAXI\,J1828--249 in the ultraviolet (with the UVOT
telescope, Fig.\,1{\it c}), its accurate position was determined
faster and easier than is usually done, when the identification
of a transient in a crowded OIR field requires a meticulous
analysis of the variability of all the infrared sources falling
within it (Fig.\,1{\it d}). 
 
Figure\,2 shows the light curve of the source constructed
in four successive energy bands in the range
from 2 to 60 keV from the MAXI, SWIFT, and INTEGRAL
data. The light curve spans the period from
the discovery of the transient (on October 15, 2013)
to the end of February 2014. In the soft 2--4 keV X-ray
band, the outburst of the transient has a FRED
shape with a fast ($\sim5$ days) rise to $\sim 100$--$150$ mCrab
and a slow ($\sim50$ days) exponential decay typical of
X-ray novae. The second, much fainter outburst
of the source began $\sim50$ days later, which is also
typical of novae (the so-called ``knee'' in their light
curves). Unfortunately, 5--10 days after the onset of
the second outburst, the sky region with the source
was no longer observable; therefore, we know little
about the second outburst. After the resumption
of observations in January 2014, the flux in soft X-ray
bands (especially in the 4--10 keV band) was
still appreciably higher than that before the onset of
the second outburst. Note the short ($\sim10$ days) dip by $\sim
40-50$\% in the 2--4 keV light curve observed
shortly after maximum light. A similar dip was observed
previously in the light curve of another X-ray
nova, MAXI\,J1836-194 (Grebenev et al. 2013),
and was explained by the transition of the source to
a harder spectral state in the standard X-ray band,
which is probably associated with the disappearance
of the soft blackbody spectral component (or a noticeable
decrease in its temperature).

\begin{figure}[tp]

\epsfxsize=0.88\textwidth
\epsffile{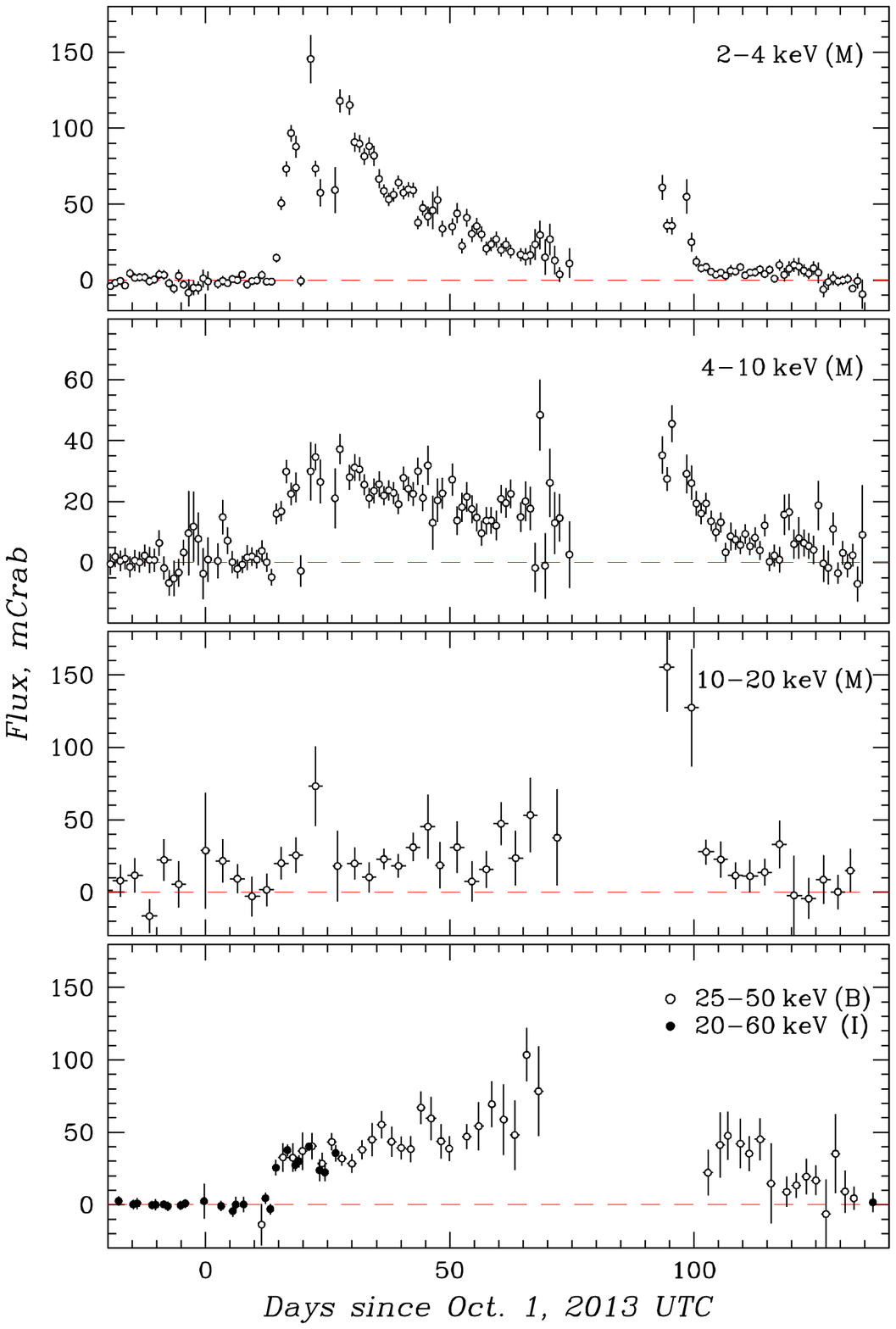}

\caption{\rm X-ray light curve of MAXI\,J1828--294 from its
  discovery on October 15, 2013, to the end of February 2014
  from the MAXI, SWIFT, and INTEGRAL data in different energy
  bands. The open circles represent the GSC/MAXI data (M) on the
  upper three panels and the BAT/SWIFT data (B) on the lower
  panel. The filled circles on the lower panel indicate the
  ISGRI/INTEGRAL data (I). The MAXI data points were obtained
  during $\sim1$ day of observations at energies $<10$ keV and three
  days at energies 10--20 keV; the ISGRI and BAT data points were
  obtained during 1 and 2 days, respectively; the actual MAXI
  and BAT exposure time often did not exceed a couple of
  hours.}
\end{figure}

In the hard ($\ga20$ keV) X-ray band, the flux from the source
after its sharp rise on October 15, 2013, to $\sim40$ mCrab
subsequently changed gradually, showing a slow rise to $\sim60$
mCrab within the first two months after the outburst and a
succeeding equally slow decay to $\sim40$ mCrab. The flux
dropped abruptly by a factor of $\sim4$ in $\sim100$ days after
the onset of the outburst and then continued to slowly
decrease. Note that the SWIFT/XRT spectral measurements
performed at this time (on February 14, 2014) showed that the
source passed to a hard state and its X-ray 0.5--10 keV spectrum
was described by a simple power law with a photon index of
$1.7\pm0.15$ (Tomsick and Corbel 2014). The first detection of
radio emission from \mbox{MAXI\,J1828--294} (Corbel 2014) also
refers to this time. The radio source was detected with a flux
density of $\simeq1.3$ mJy at 3.5 cm and had a flat spectrum
suggesting a synchrotron origin and self-absorption.

\begin{figure}[t]\centering
\epsfxsize=\textwidth
\hspace{1cm}
\epsffile{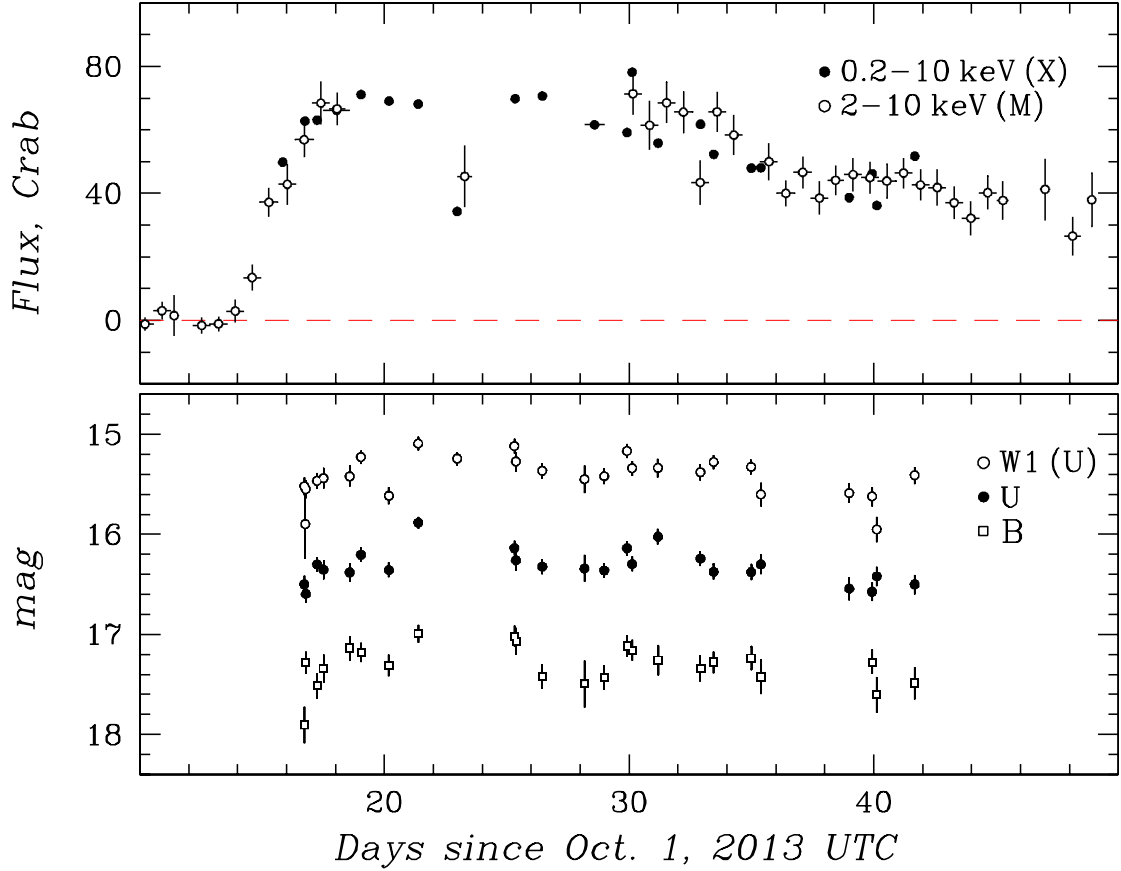}
\caption{\rm Comparison of the soft X-ray ($<10$ keV) and OUV
  light curves for the initial stage of the outburst of
  MAXI\,J1828--249 (October --- November 2013). The filled and open
  circles on the upper panel indicate the XRT/SWIFT (X) and GSC/MAXI (M)
  X-ray flux measurements, respectively; the UVOT/SWIFT (U) flux
  measurements in the W1, U, and B filters are shown on the lower
  panel. The W1 light curves is shifted by 1.2 mag upward for
  clarity. The resolution corresponds to about 18--24 h.}
\end{figure}
\begin{figure}[t]\centering
\epsfxsize=\textwidth
\hspace{1cm}
\epsffile{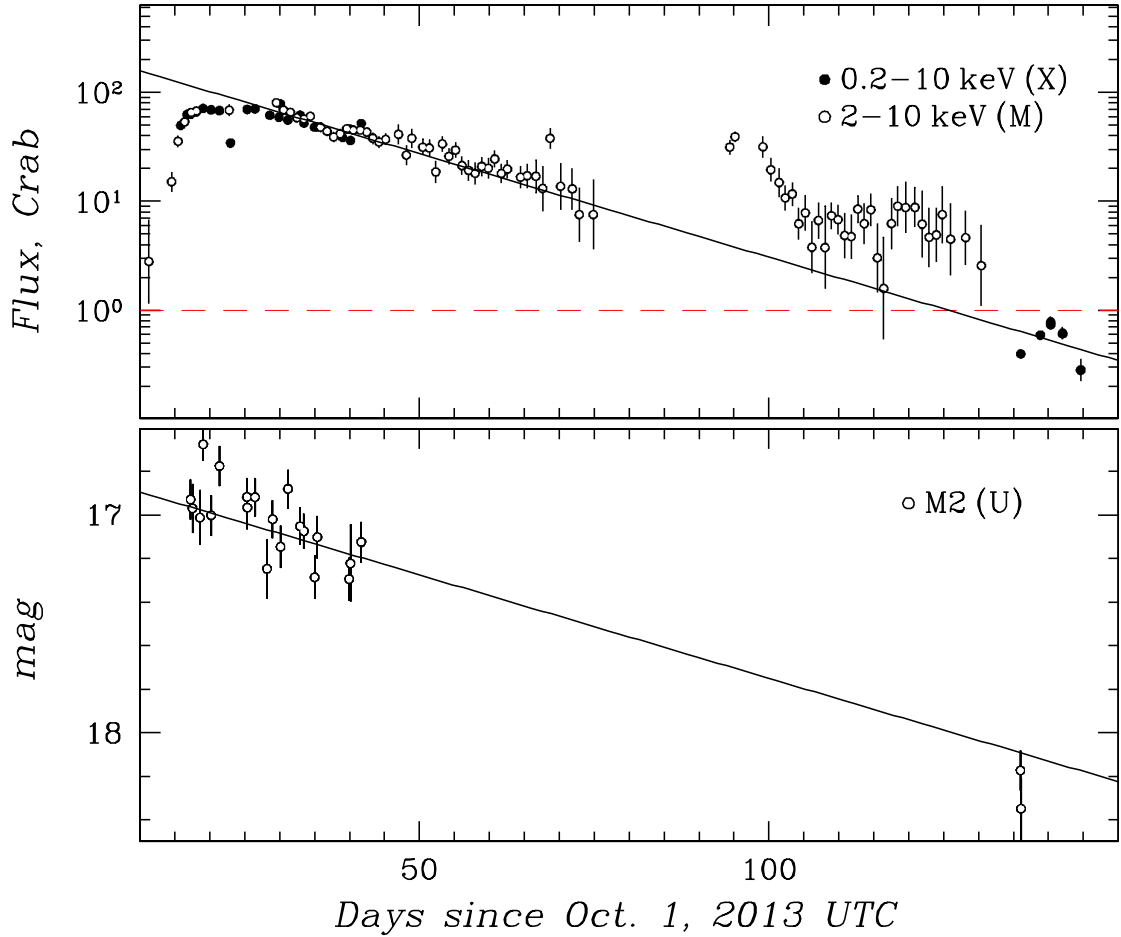}
\caption{\rm Comparison of the long-term (October 2013 ---
  February 2014) X-ray ($<10$ keV) and optical (M2 filter) light
  curves for MAXI\,J1828--249 from the data of the XRT (X) and UVOT
  (U) telescopes of the SWIFT observatory and the GSC instrument
  of the MAXI observatory (M).}
\end{figure}

The light curve in Fig.\,3 (filled circles) shows the initial
activity stage of the source in the 0.2--10 keV band. It was
obtained by the XRT/SWIFT X-ray telescope with smaller errors
and more complete coverage than those for the GSC/MAXI
instrument (open circles, the 2--10 keV band). According to this
light curve, the flux from the source reaches its maximum
already by October 18 and then began to gradually decrease.  The
figure confirms the presence of the above dip in the light curve
by $\sim50$\% near October 25. The short ($\sim1$ day) spike in
the light curve near October 30 is also of interest. The lower
panel in the figure presents the optical and ultraviolet (U, B,
and W1) light curves constructed from the UVOT/SWIFT
measurements. The observed variability follows the soft X-ray
variability in many details, suggesting that the OUV emission
closely correlates with the X-ray one, i.e., it originates in
the same region or is the result of its direct reprocessing
under accretion disk irradiation. This can be seen even better
from Fig.\,4, which presents the same X-ray light curves but
over a longer period (October 2013 --- February 2014) in
comparison with the UVOT M2 light curve. A logarithmic axis is
used for the X-ray flux in this figure; the straight lines
specify an exponential decay with the same decay time scale of
$\simeq 53$ days for the X-ray and optical bands. The flux in
both bands is seen to actually decrease according to a single
law.

\begin{figure}[t]
\epsfxsize=\textwidth
\epsffile{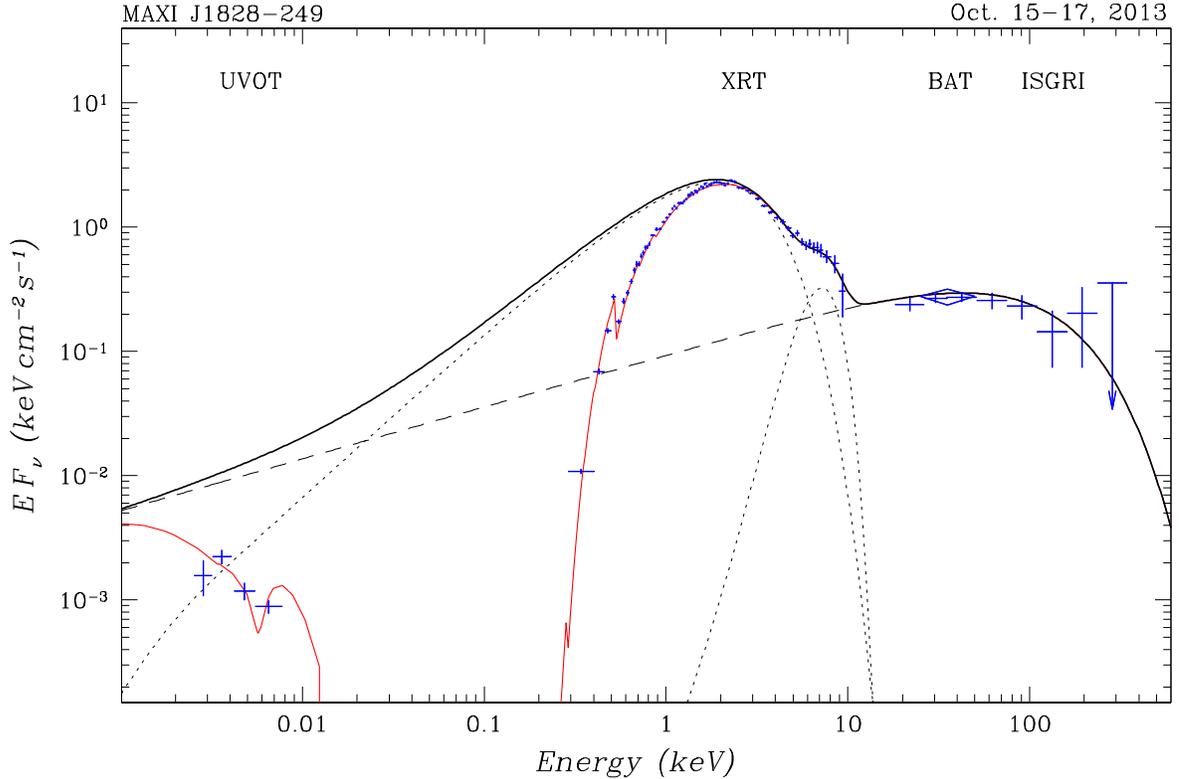}

\caption{\rm Broadband (0.002--400 keV) spectrum of
  MAXI\,J1828--249 obtained by INTEGRAL and SWIFT on October
  15--18, 2013. The thick solid (black) line indicates its fit
  by the adopted model (see the text); the thin solid (red) line
  represents the same model after the correction for
  interstellar absorption. The dashed line indicates the hard
  component of the spectral model connected with Comptonization
  (fitted by a power law with an exponential cutoff at high
  energies); the dotted lines indicate other components of the
  model: blackbody disk radiation and a Gaussian iron
  fluorescence line at 6.4 keV.}
\end{figure}
It is difficult to judge the shape and evolution of the
source's broadband spectrum from the light curves
alone. It is necessary to have the spectrum itself
at various outburst stages. To obtain such spectra,
we selected seven time intervals with maximally wide
energy coverage, from the optical to the hard X-ray
bands, from all of the available data. The start (end)
date and time of observations for the instruments
operated in each interval are given in Table\,1.

The source's spectrum taken during the first interval of
observations, on October 15--18, 2013, i.e., almost immediately
after its discovery, is presented in Fig.\,5.  We used the
quasi-simultaneous IBIS/ISGRI, BAT, XRT, and UVOT
observations. The relative data normalization was set equal to
unity for all instruments.  The thin solid line indicates the
best fit to this spectrum; the thick solid line indicates the
same model spectrum corrected for interstellar absorption, i.e.,
the original spectrum of the source. The absorption was
described by the fit from Morrison and McCammon (1981, the {\sc
  wabs} code in XSPEC) in the low-energy part of the X-ray
spectrum and by the {\sc redden} code, where the color
correction \mbox{E(B-V)} was assumed to be equal to$N_{\rm
  H}/(5.6\times10^{21} \ \mbox{\rm cm}^{-2})$ (see Draine 2003),
in the optical spectrum. Note once again that the measured
$N_{\rm H}$ turned out to be very close to the average Galactic
absorption expected in this direction, $N_{\rm H}\simeq
(1.7\pm0.2) \times10^{21}\ \mbox{\rm cm}^{-2}$ (Kalberla et
al. 2005).

\begin{table}[t]
  \small
\centering {{\bf Table 1.} Measurements of the spectrum
  for \mbox{MAXI\,J1828--249}  \protect\\ at various stages of its outburst\a\protect\\}

\vspace{5mm}

\begin{tabular}{l|r|r|r|r|c} \hline\hline
\multicolumn{1}{c|}{Date}& \multicolumn{2}{c|}{OIR \& UV} & 0.4--10 keV & \multicolumn{2}{c}{$>20$ keV}\\ \hline
         & RTT-150& UVOT &  \multicolumn{1}{c|}{XRT}& \multicolumn{1}{c|}{BAT} &IBIS \\ \hline 
\multicolumn{6}{c}{2013}\\ \hline
    Oct. 15--18&  &17 17\uh15\um&  \multicolumn{1}{c|}{--$\,^{\prime\prime}$--}& \multicolumn{1}{c|}{--$\,^{\prime\prime}$--} &15 18\uh03\um--18 02\uh58\um\\
     Oct. 18--20&  &20 00\uh55\um&  \multicolumn{1}{c|}{--$\,^{\prime\prime}$--}& \multicolumn{1}{c|}{--$\,^{\prime\prime}$--}&18 21\uh55\um--20 09\uh13\um\\ 
     Oct.  21--22&  &21  04\uh26\um&  \multicolumn{1}{c|}{--$\,^{\prime\prime}$--}& \multicolumn{1}{c|}{--$\,^{\prime\prime}$--}&21 17\uh35\um--22 13\uh39\um\\  
 Nov. 9& 15\uh42\um&22\uh14\um &\multicolumn{1}{c|}{--$\,^{\prime\prime}$--}   & \multicolumn{1}{c|}{--$\,^{\prime\prime}$--}&\\ 
 Nov. 10& 15\uh52\um&02\uh50\um & \multicolumn{1}{c|}{--$\,^{\prime\prime}$--}  & \multicolumn{1}{c|}{--$\,^{\prime\prime}$--}&\\ 
 Nov. 11--12&12  15\uh45\um& 11 16\uh01\um & \multicolumn{1}{c|}{--$\,^{\prime\prime}$--}& \multicolumn{1}{c|}{--$\,^{\prime\prime}$--}&\\ \hline 
\multicolumn{6}{c}{2014}\\ \hline
Feb. 13--14&&13  23\uh59\um
&\multicolumn{1}{c|}{--$\,^{\prime\prime}$--}&\multicolumn{1}{c|}{--$\,^{\prime\prime}$--}
& 14  13\uh45\um--17\uh26\um\\ \hline
\multicolumn{6}{l}{}\\ [-3mm]
\multicolumn{6}{l}{\a\ The start (end) date and time (UT) of the
  interval of observations. }\\
\multicolumn{6}{l}{\ \  \ The SWIFT and RTT-150 observations
  usually lasted $\la1000$ s.}\\ [0mm]
\end{tabular}
\end{table}

The dashed and dotted lines in the figure indicate the
individual components of the spectral model used: (1) multicolor
blackbody disk radiation ({\sc diskbb} in the XSPEC code;
Shakura and Sunyaev 1973), (2) a power law with an exponential
cutoff at high energies ({\sc cutoffpl} in the XSPEC code), and
(3) a Gaussian emission line to describe the iron fluorescence
at 6.4 keV ({\sc gaussian}). Note that the {\sc diskbb} model
takes into account only the energy release due to viscous energy
dissipation in the disk.The Gaussian line can also take into
account possible excess emission formed in the transition region
between the cold disk and the high temperature central
cloud. The Gaussian line width was fixed at $\sigma=1.4$ keV,
which roughly corresponds to the detector resolution. Other
best-fit parameters and the corresponding luminosities are given
in Table\,2. When estimating the luminosity, we assumed
\mbox{MAXI\,J1828--249} to be near the Galactic center at a
distance $d\simeq8$ kpc.

The spectrum of the blackbody disk presented in Fig.\,5 was
obtained by the integration over its surface from $R_1=3R_{\rm
  g}\simeq9\times10^6\ (M/10\ M_{\odot})\ \mbox{cm} (=R_0)$ to
$R_2=1.5\times10^5 R_{\rm g}\simeq 4.5\times10^{11}
(M/10\ M_{\odot})\ \mbox{cm}$ and, thus, was computed more
accurately than in the {\sc diskbb} function\footnote{In the
  {\sc diskbb} function of the XSPEC code, the integration is
  from the radius corresponding to the maximum disk temperature
  to infinity; therefore, if the blackbody disk extends
  sufficiently close to the center, to (or almost to) the radius
  of the innermost stable orbit $R_0=3\, R_{\rm g}$, then the
  radiation from the innermost region, where the surface
  temperature already drops below the maximum one, turns out to
  be disregarded.}. Such a spectrum describes the soft component
in the X-ray spectrum appreciably better than is done by the
{\sc diskbb} function. Here and below, $R_{\rm g}=2GM/c^2$ is
the gravitational radius of the black hole and
$M\simeq10\ M_{\odot}$ is its mass. The chosen outer disk radius
$R_2$ corresponds to the separation between the components of
the binary system with a period of $\sim10$ h, typical of many
X-ray novae (see, e.g., Cherepashchuk 2013). Restricting the
disk size leads to a cutoff of its blackbody spectrum in the far
infrared (Fig.\,5).

\begin{table}[t]
\small
\centering{{\bf Table 2.} Fits to the radiation spectrum of \mbox{MAXI\,J1828--249}\protect\\} 

\vspace{2mm}

\hspace{-4mm}\begin{tabular}{@{}c@{}|r@{}c@{}l|r@{}c@{}l|r@{}c@{}l|r@{}c@{}l|r@{}c@{}l|r@{}c@{}l|r@{}c@{}l@{\hspace{0.5mm}}} \hline\hline
Model&\multicolumn{21}{c}{Date}\\ \cline{2-22}
&\multicolumn{3}{c|}{Oct.\,15--18,}
&\multicolumn{3}{c|}{Oct.\,18--20,}
&\multicolumn{3}{c|}{Oct.\,21--22,}
&\multicolumn{3}{c|}{Nov.\,9,}
&\multicolumn{3}{c|}{Nov.\,10,}
&\multicolumn{3}{c|}{Nov.\,11--12,}
&\multicolumn{3}{c}{Feb.\,13,}\\
&\multicolumn{3}{c|}{2013}
&\multicolumn{3}{c|}{2013}
&\multicolumn{3}{c|}{2013}
&\multicolumn{3}{c|}{2013}
&\multicolumn{3}{c|}{2013}
&\multicolumn{3}{c|}{2013}
&\multicolumn{3}{c}{2014}\\ \hline
\multicolumn{22}{c}{Power-law model with exponential cutoff}\\ \hline
$N_{\rm H}$\a &~2.05&$\pm$&0.03 &~2.11&$\pm$&0.03 & ~1.93&$\pm$&0.03 &
                         2.00&$\pm$&0.03 & 2.09&$\pm$&0.03 &~1.93&$\pm$&0.03 &
                         2.25&$\pm$&0.24\\

$\alpha$\b      &1.60&$\pm$&0.02 & 1.69&$\pm$&0.01 & 1.63&$\pm$&0.01 &
                         1.42&$\pm$&0.01 & 1.44&$\pm$&0.01 & 1.43&$\pm$&0.01 &
                         1.73&$\pm$&0.05\\
                         
$E_{\rm br}$\c &108&$\pm$&27     &  139&$\pm$&41    & 93&$\pm$&21 &
                       \multicolumn{3}{c|}{20}& \multicolumn{3}{c|}{20}& 17.3&$\pm$&2.8 &
                        \multicolumn{3}{c}{>150}\\
                         
$I_0$\d           &0.84&$\pm$&0.06 & 1.02&$\pm$&0.06 & 1.08&$\pm$&0.07 &
                        2.28&$\pm$&0.08 &2.05&$\pm$&0.07  & 2.05&$\pm$&0.12 &
                        0.19&$\pm$&0.01\\

$kT_{bb}$\e    &779&$\pm$&5 & 822&$\pm$&4 & 775&$\pm$&5&
                         659&$\pm$&6 & 685&$\pm$&6 & 681&$\pm$&6 &
                        \multicolumn{3}{c}{-}\\

$R_{\rm  bb}\sqrt{cos\,i}$\,\f{$\,$}&24.8&$\pm$&0.3 & 24.9&$\pm$&0.3 & 28.4&$\pm$&0.3&
                                             27.1&$\pm$&0.5 & 26.3&$\pm$&0.5 & 27.3&$\pm$&0.4&
                                             \multicolumn{3}{c}{-}\\

$\xi_{\rm RTT}$\g\ &\multicolumn{3}{c|}{-}&\multicolumn{3}{c|}{-}&\multicolumn{3}{c|}{-}&
       0.79&$\pm$&0.06 & 0.80&$\pm$&0.05 & 0.88&$\pm$&0.05&
       \multicolumn{3}{c}{-}\\ 

$I_{6.4}$\h\ &15.5&$\pm$&2.2 & 6.6&$\pm$&2.0 & 5.2&$\pm$&2.4 &
       \multicolumn{3}{c|}{-}&\multicolumn{3}{c|}{-}&\multicolumn{3}{c|}{-}&
       \multicolumn{3}{c}{-}\\ 

$L_{\rm  X}^{\rm S~}$\i\ &72.5&$\pm$&5.6 & 100.6&$\pm$&5.9 & 94.3&$\pm$&6.5&
       52.3&$\pm$&1.8 &56.8&$\pm$&2.1&58.6&$\pm$&3.6&6.97&$\pm$&0.51\\ 

$L_{\rm  X}^{\rm H1~}$\i\ &6.68&$\pm$&0.52 & 6.80 &$\pm$&0.40
       &6.97&$\pm$&0.40&\multicolumn{3}{c|}{-} &\multicolumn{3}{c|}{-}
       &\multicolumn{3}{c|}{-} & 1.81&$\pm$&0.13\\
       
$L_{\rm  X}^{\rm H2~}$\i\ &2.29&$\pm$&0.18 & 2.20 &$\pm$&0.13 & 2.53&$\pm$&0.17 &       3.01&$\pm$&0.11 & 2.46&$\pm$&0.09& 2.00&$\pm$&0.12& 0.43&$\pm$&0.03\\
       \hline
\multicolumn{18}{l}{}\\ [-3mm]
\multicolumn{18}{l}{~\a\ Hydrogen column density, $10^{21} \ \mbox{cm}^{-2}$.}\\
\multicolumn{18}{l}{~\b\ Photon index.}\\
\multicolumn{18}{l}{~\c\ Energy of the high energy cutoff, keV.}\\
\multicolumn{18}{l}{~\d\ Normalization of the power-law spectral
  component, $10^{-1}$ phot cm$^{-2}$ s$^{-1}$ keV$^{-1}$.} \\
\multicolumn{18}{l}{~\e\ Temperature of the disk at its inner edge, eV.}\\
\multicolumn{18}{l}{~\f\  ~Inner radius of the disk assuming  $d=8$ kpc
  ($i$ --- disk inclination), km.}\\
\multicolumn{18}{l}{~\g\ Relative normalization of the RTT-150 telescope data.}\\
\multicolumn{18}{l}{~\h\ Intensity of the 6.4 keV line,
  $10^{-3}$ phot cm$^{-2}$ s$^{-1}$.}\\ 
\multicolumn{18}{l}{~\i\ Luminosity in the energy ranges 0.001--13 (S), 13--400 (H1) and 25--50 (H2) keV}\\
\multicolumn{18}{l}{~~ assuming $d=8$ kpc  after correction for photoabsorption,
   $10^{36} \ \mbox{erg s}^{-1}$.}
\end{tabular}
\end{table}

It can be seen from Fig.\,5 that at energies below $\sim25$ eV,
the observed spectrum of the source within this model is
dominated by a power-law emission component that exceeds the
flux due to viscous energy dissipation in the outer disk regions
by almost an order of magnitude. It is believed that the X-ray
emission from the inner disk regions can be intercepted by the
outer regions and, being reprocessed in them, can raise the disk
surface temperature. The OIR emission from the system must
increase accordingly. This effect should be included in our
analysis. The flux reradiated by a disk surface unit in the
infrared and optical ranges is given by the equation (Shakura
and Sunyaev 1973; Lyuty and Sunyaev 1976)

\begin{equation}\label{e10}
Q_{\rm irr}=\frac{L_{\rm
X}\,(1-\beta_d)}{4\,\pi\,R^2}\,\left(\frac{H}{R}
\right)^m\,\left(\frac{d\ln{H}}{d\ln{R}}-1\right).
\end{equation}

Here, $H\sim R^{\,\gamma}$ is the disk half-thickness at a given
radius, $\beta_d\simeq0.9$ is the X-ray albedo of its surface
(which take into account the fact that the emission
is incident at very small angles; see, e.g., de
Jong 1996), $m = 1$ if the inner disk region is puffed up
by the thermal and secular instabilities and, therefore,
radiates almost isotropically and $m = 2$ if the emitting
region has a flat surface. We will assume $m = 1$, because
the X-ray heating obviously turns out to be insignificant at $m =
2$. Comparison of  $Q_{\rm irr}$ with the flux due to 
viscous energy dissipation in the disk (Shakura and
Sunyaev 1973)
$$Q_{\rm
  vis}=\frac{3}{8\,\pi}\,\frac{GM\dot{M}}{R^{3}}\,\left[1-
  \left(\frac{R_0}{R}\right)^{1/2}\right]\simeq
\frac{3\,L_{d}}{4\pi\,R^2} \left(\frac{R_0}{R}\right)$$ shows
that the X-ray heating of the disk surface dominates over the
viscous one at distances $R$ from the center exceeding the
critical radius $$R_{\rm irr}\simeq
\frac{30}{(\gamma-1)}\,\left(\frac{L_d}{L_{\rm X}}\right)\,
\left(\frac{H}{R}\right)^{-m} \,R_0\sim36000 (L_d /L_{\rm X})
R_0.$$ The regions near $R_{\rm V}\simeq 9600
(L_d\,/10^{38}\ \mbox{erg s}^{-1})^{1/3} R_0< R_{\rm irr}$
mainly contribute to the optical (V filter) disk emission. Here,
$L_{d}\simeq0.08\dot{M}c^2\gg L_{\rm X}$ is the total energy
release in the disk due to turbulent viscosity. It is clearly
seen from Fig.\,5 that the total luminosity of the hard
power-law component ($\simeq L_{\rm X}$) is exceeded
considerably, by a factor of 5--10, even by the luminosity of
the blackbody accretion disk, which is lower than $L_{d}$. In
the standard model of Shakura and Sunyaev (1973) $H/R\simeq
6.7\times10^{-3} (R/R_0)^{1/8}$ ($\gamma=9/8$ ) depends weakly
on $R$.  The disk surface temperature in the heated region
$T_{\rm s}=(Q_{\rm irr}/\sigma+Q_{\rm
  vis}/\sigma)^{1/4}\simeq(Q_{\rm irr}/\sigma)^{1/4},$ where 
$\sigma$ is the Stefan-Boltzmann constant, decreases with radius
as $\sim R^{-1/2+m/32}\sim R^{-15/32}$ (see Eq. \ref{e10}). In
the disk region where the influence of surface heating is minor,
the temperature decreases faster, $T_{\rm s}\sim R^{-3/4}$
(Shakura and Sunyaev 1973).

\begin{figure}[t]
\epsfxsize=\textwidth
\epsffile{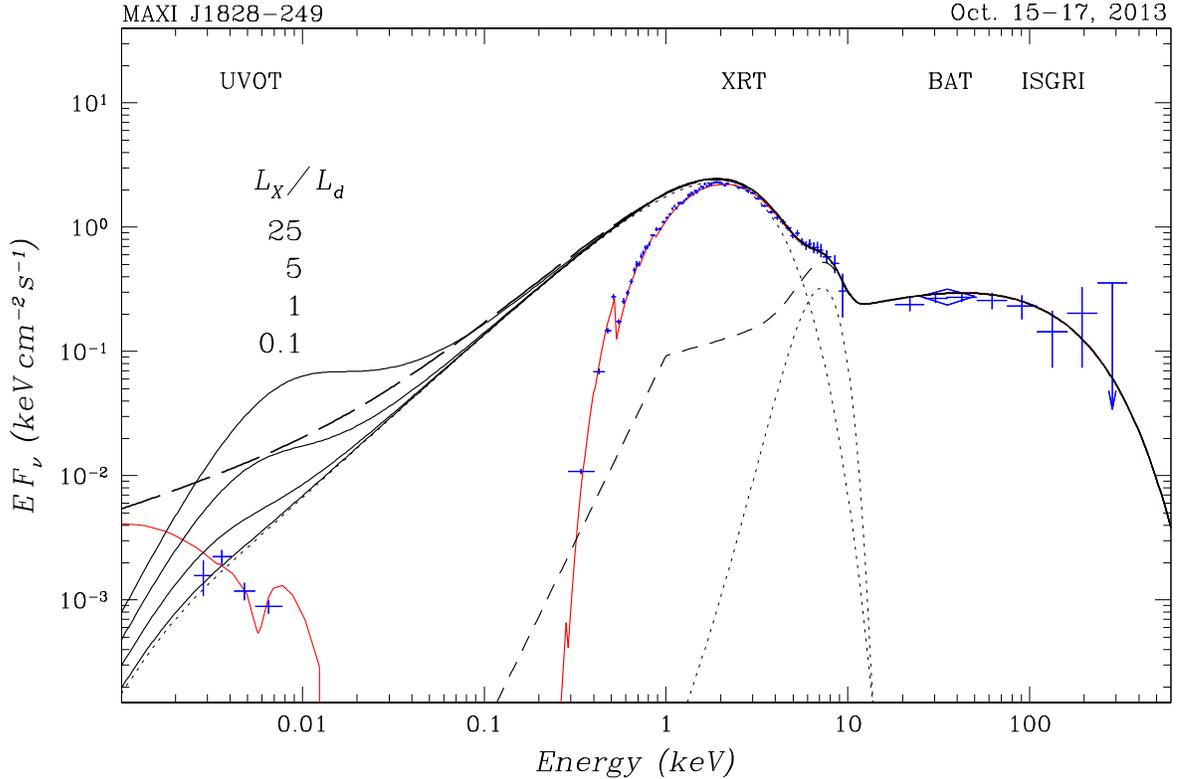}

\caption{\rm The same broadband spectrum of MAXI\,J1828--249
  obtained by INTEGRAL and SWIFT on October 15--18, 2013, as
  that in Fig.\,5 but fitted by the model with a power-law
  component disappearing fast below 1 keV (dashed line with
  short dashes) that takes into account the irradiation of the
  multicolor blackbody accretion disk by hard emission from the
  central disk region (solid lines). Various possible ratios of
  the hard X-ray luminosity to the total disk luminosity,
  $L_{\rm X}/L_{d}=25, 5, 1, 0.1$, are considered.  The dashed
  line with long dashes indicates the fit with the dominant
  contribution of a continuous power-law component (see
  Fig.\,5).}
\end{figure}
\begin{figure}[t]
\epsfxsize=\textwidth
\epsffile{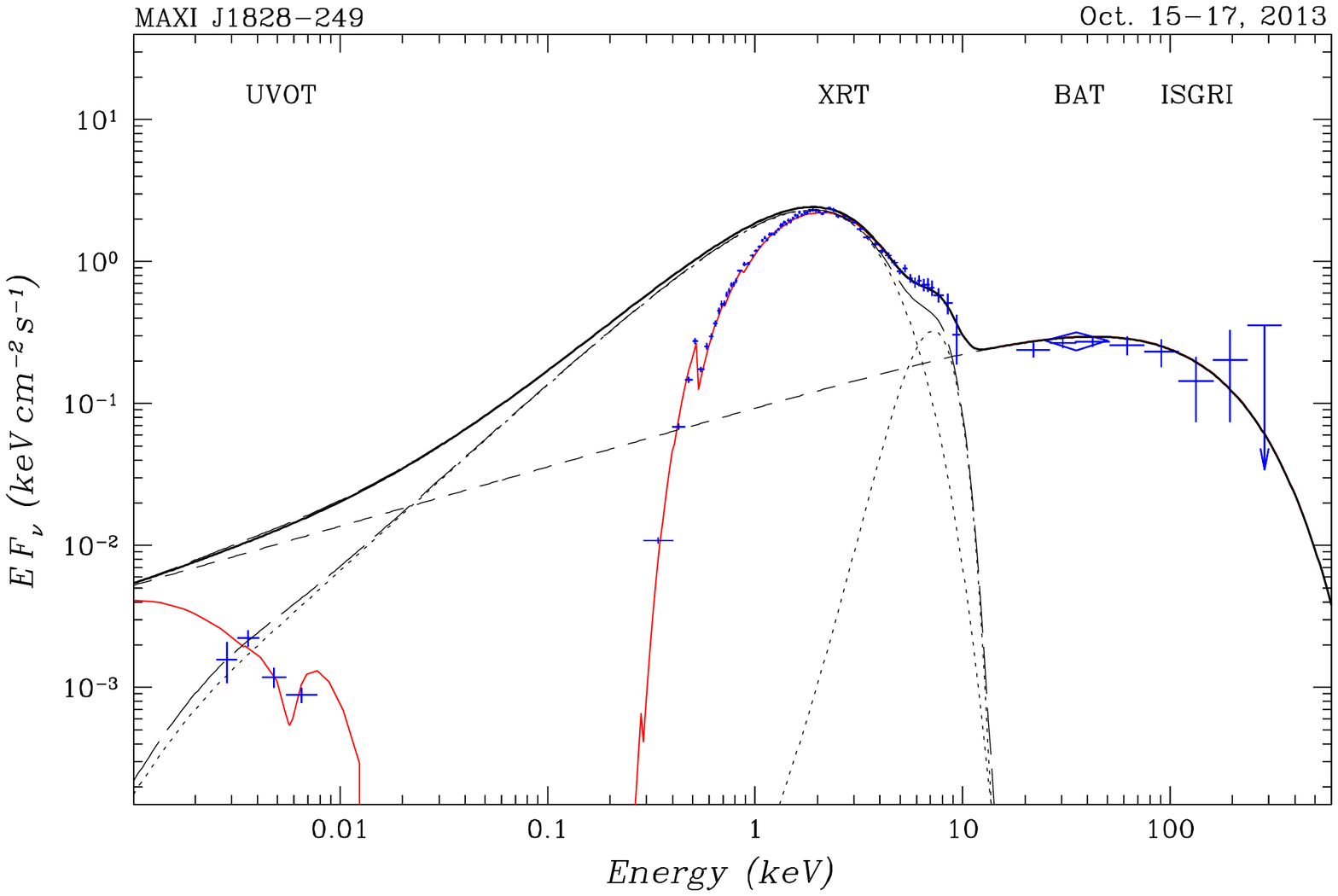}

\caption{\rm The same broadband spectrum of MAXI\, J1828--249
  obtained by INTEGRAL and SWIFT on October 15--18, 2013, as
  that in Fig.\,5 but fitted by the model that takes into account
  the irradiation of the multicolor blackbody accretion disk by
  hard emission from its central region at  $L_{\rm X}\simeq
  0.25\ L_{d}$ (long dashes).} 
\end{figure}

The solid lines in Fig.\,6 indicate the spectra of such an
X-ray-heated disk obtained for various $L_{\rm X}/L_{d}$ ratios
by assuming the power-law component to undergo a break in the
hard X-ray band (near $\sim1$ keV) and drops according to the
Rayleigh-Jeans law at lower energies (virtually without
contributing to the optical and infrared emissions from the
system).  From comparison with the power-law spectral component
in Fig. 5 shown in this figure by long dashes we see that
reradiation (X-ray disk surface heating) could actually explain
the observed OIR emission from MAXI\,J1828--249, but only under
the condition that $L_{\rm X} \ga 5 L_{d}$, i.e., in the
presence of an additional intense central X-ray source. In the
case of accretion onto a black hole, only the emission from the
jets could be such a source that is scarcely probable.
Figure 7 shows the spectrum
with a dominant (in the optical band) power-law
component, as in Fig.\,5, but with a moderate (more
realistic) contribution of the radiation from the heated
disk, in accordance with $L_{\rm X}=0.25 L_{d}$. Although the
heating raises appreciably the optical and infrared
disk luminosities, it is clearly insufficient to explain
the observed optical emission from the system. Besides,
the specific hard X-ray luminosity of this source
during the observations being discussed was actually
much lower than this value, as is clearly seen from
comparison of the amplitudes of the hard and soft
(blackbody) X-ray spectral components presented in
Fig.\,5 (the spectra are given in the form of  $E\,F_{\nu},$ where
$F_{\nu}$ is the radiation spectrum; therefore, their amplitudes
are proportional to the luminosity in the logarithmic
scale of the figure) and the estimates of the source's luminosity
in the soft and hard bands presented in Table 2. 

All of the broadband spectra measured near or immediately after
the maximum light of the X-ray nova MAXI\,J1828--249
(corresponding to the first six intervals of observations) can
be fitted in a similar way. The parameters of the spectra are
given in Table 2. The increase in the slope of the power-law
component during the first week of observations turned out to be
not so dramatic as was suggested by Krivonos and Tsygankov
(2013). In subsequent days the slope even slightly decreased
(till $\sim1.4$) but returned to the initial value of $\sim1.7$ by
February. Of course, the real hardness of the spectrum in the
X-ray band depended not only on this slope but also on energy of
the exponential cutoff decreasing with time. The location of the
inner edge of the blackbody region of the disk established shortly
after the outburst onset subsequently barely changed, while the
temperature of the disk surface gradually decreased. The
intensity of the iron fluorescent line also decreased rapidly;
it was noticeable only within the first days after the outburst
when the disk approached the black hole quite closely. Three
spectra corresponding to different phases of the source's
evolution are shown in Fig. 8 in comparison with the model
spectrum obtained during the first observation. The third
spectrum shown in this figure was taken in February 2014 at the
decaying phase of the outburst (the seventh time interval in
Table 1). This spectrum differs noticeably from the first six
spectra, being a purely power-law one, without any signatures of
the blackbody component.
\begin{figure}[ht]
\epsfxsize=\textwidth
  
\epsffile{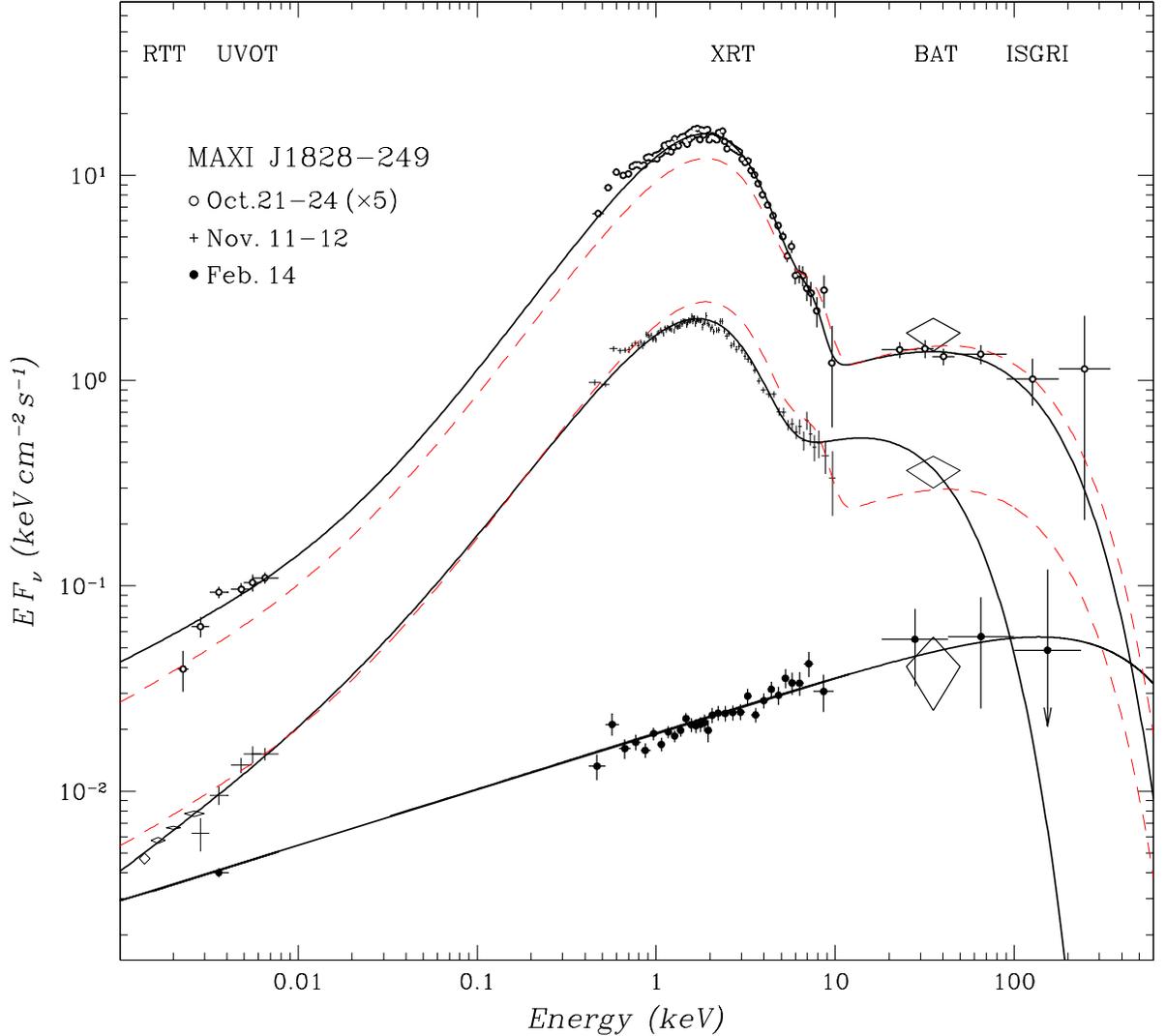}

\caption{\rm Evolution of the broadband emission spectrum of the
  X-ray nova MAXI\,J1828--249 in October 2013 --- February 2014
  according to data from INTEGRAL (the ISGRI instrument) and
  SWIFT (BAT, XRT, and UVOT) observatories, and the RTT-150
  telescope. The best-fit parameters are given in Table 2. The
  spectrum measured on October 21--24 is shifted for convenience
  upward by multiplying it by 5. For comparison, the dashed
  lines indicate the best-fit to the spectrum obtained during the
  first observation (on October 15--17, 2013).}
\end{figure}

\section*{CONCLUSIONS}
\noindent
The results of our analysis of the INTEGRAL,
SWIFT, MAXI, and RTT-150 observational data for
the X-ray nova MAXI\,J1828--249 can be summarized
as follows:
\begin{enumerate}
\item The ultraviolet, optical, and infrared emissions from the
  source during its 2013 outburst cannot be explained
  exclusively by the blackbody radiation from the distant outer
  accretion disk regions due to internal energy release and/or
  disk surface heating by hard photons emitted near the black
  hole. The blackbody radiation was weak both during the hard
  state of the source observed at the decaying phase of the
  outburst and during its soft (more likely two-component) state
  near the maximum of the X-ray flux from the nova.

\item The missing OIR emission from the source can be
  successfully explained by extrapolating to this spectral range
  the power-law emission component responsible for the hard
  power-law ``tail'' with an exponential cutoff at energies
  above $\sim100$ keV observed in the source's spectrum.

\item The optical variability of the source as a whole
coincided with the X-ray one on a time scale of days
and weeks, confirming the conclusion about a unified
spectrum. 
\end{enumerate}

\section*{ACKNOWLEDGMENTS}

This work is based on the data from the INTEGRAL observatory
retrieved through its Russian and European Science Data Centers,
the SWIFT observatory retrieved through NASA/HEASARC, the MAXI
experiment onboard the ISS retrieved through the site {\sc 
  maxi.riken.jp/top}, and the 1.5-m RTT-150 telescope. We take
the opportunity to thank the TUBITAK National Observatory
(Turkey), the Space Research Institute of the Russian Academy of
Sciences, and the Kazan Federal University for their support in
using this telescope.

This study was financially supported by the Program
of the President of the Russian Federation
for Support of Leading Scientific Schools (project
NSh-6137.2014.2). S.A. Grebenev is grateful to the Russian Foundation for
Basic Research (project no. 13-02-01375-a) for its support,
A.V. Mescheryakov is grateful for the support to the State
Program for Increasing the Competitiveness of the Kazan
Federal University. 

\pagebreak   

\begin{flushright}
{\it  Translated by V. Astakhov}
\end{flushright}
\end{document}